# A cold electron emission spectrum in a low-frequency laser field

E A Mikhin[1], A A Drobyshev[1] and P A Golovinski [1,2]

[1]Physics Research Laboratory, Voronezh State Technical University, Voronezh, 20-Letiya Oktyabrya st., 84, 394006, Russia
[2]Moscow Physical-Technical Institute (State University), Moscow region, Dolgoprudny, Institutskii per., 9, 141700, Russia

E-mail: mihinzheny@mail.ru

**Abstract**. We consider the process of cold emission of electrons from a metal in a strong low-frequency laser field. An expression is obtained for the electron spectrum with allowance for their rescattering at the metal surface as a result of an extension of the Fowler-Nordheim theory and modification of the Simplemen model. The numerically simulated features of the electron spectrum are similar to the ATI spectra for atoms and have the same form as the experimental surface emission dependencies for a metal.

## 1. The emission current density in a low-frequency field

The steady interest of researchers to the process of electron emission from the metal surfaces in the laser field is due both to the improvement of laser technologies that make it possible to reliably control the parameters of the laser field and also the development a technology of nanoscale structures. The combination of these factors determines the area of laser applications associated with the observation, control and diagnostics of nanoscale objects, including a cold electron emission in femtosecond laser fields [1,2].

In a seminal field emission study, Fowler and Nordheim [3] have obtained an expression for the electron current density

$$J(t) = \frac{1}{16\pi^2} \frac{F^2}{W t^2(y)} \exp\left(-\frac{4(2W^3)^{1/2}}{3F} v(y)\right). \quad (1)$$

Hereinafter, we use the atomic system of units where $e = m = \hbar = 1$. Equation (1) demonstrates the dependence of the current density on the work function $W$ of the electron from the metal and on the strength of the electric field $F$. The function $t(y)$ in equation (1) is close to unity and varies little with the argument $y = F^{1/2}/W$ changing, and the slowly varying function $v(y)$ is called the Nordheim function. With a high accuracy, we can use a simple approximation [4]

$$v(y) = 1 - y^2 + y^2 \ln y / 3. \quad (2)$$

The Fowler-Nordheim field emission model is based on a number of assumptions. The metal is described in terms of the Sommerfeld theory, i.e. electrons within the metal are assumed to be free. On the boundary of a metal with a vacuum, the potential varies abruptly and outside the metal it is composed of the image potential and the potential of the electrostatic field

$$U(x) = \begin{cases} -U_0, & x \leq 0, \\ -\dfrac{1}{4x} - Fx, & x > 0. \end{cases} \quad (3)$$

The constant $U_0 > 0$ determines the position of the bottom of the potential well inside the metal. The coordinate axis $OX$ is perpendicular to the surface of the metal and is directed from the surface to the side of the vacuum. The one-dimensionality of the potential in equation (3) means that we neglect inhomogeneity and defects in the structure of the emitter. If the metal is at room temperature, this temperature is small compared to the Fermi temperature, which is 60 000 K at the Fermi energy $E_F = 5$ eV. Then thermal emission does not lead to a significant change in the emission current density, and there are no temperature factors in equation (1).

The current of cold electron emission in vacuum in a low-frequency laser field is caused by the tunneling of electrons from the metal through the barrier defined by equation (3). The flux of emitted electrons at each instant of time can be fined as the number of collisions of metal electrons with the surface multiplied by the transmission coefficient, the value of which is determined by the quasiclassical WKB method. Following these assumptions, we consider the emission of electrons in an external low-frequency field

$$F(t) = -F_0 \sin(\omega t), \qquad (4)$$

where $F_0$ is the amplitude value of the field strength, $\omega$ is the field frequency. The initial phase is chosen so that in the first half-period the electron experiences acceleration in the positive direction of the axis, i.e. electronic emission is possible. The low frequency of the field lies in the smallness of the laser frequency $\omega$ compared with the characteristic tunneling frequency $\omega_t = F_0/\sqrt{2W}$, which is equivalent to the smallness of the Keldysh adiabaticity parameter $\gamma = \omega/\omega_t \ll 1$. Under these conditions, the field does not have time to change significantly during the tunneling of an electron through a potential barrier, and the process can be regarded as occurring in a quasistationary field with a strength $F(t)$. The emission current becomes a function of time and its density can be obtained in accordance with equation (1), by replacement $F \to |F(t)|$ in the first half-period of the field. In the second half-cycle, there is no electron emission. Further, all the relations that are satisfied for the first period of the field will be valid for the remaining periods, and therefore, in obtaining electron spectra, we confine ourselves to considering emission in one half-cycle $T/2$ of the laser field.

**2. The spectrum of direct emission electrons**

In tunnel emission in a low-frequency field, electrons appear after the barrier with zero initial velocity at different instants of time $t_0$ and at different phases of the field $\varphi = \omega t_0$ [5]. To determine their energy in the final state, we use the classical equation of motion

$$\ddot{x} = F_0 \sin(\omega t) \qquad (5)$$

with the boundary conditions $x(t_0) = 0$, $\dot{x}(t_0) = 0$. After a single integration of equation (5), we obtain the electron momentum

$$p(t) = \frac{F_0}{\omega}(\cos(\omega t_0) - \cos(\omega t)). \qquad (6)$$

The average over the period of the field depends on the initial phase $\varphi$ of the field, and we have

$$\overline{p} = \frac{F_0}{\omega}\cos(\omega t_0). \qquad (7)$$

The energy of the electron in the final state is

$$E = 2\frac{F_0^2}{4\omega^2}\cos^2(\omega t_0) = 2U_p \cos^2(\omega t_0), \qquad (8)$$

where $U_p$ is the ponderomotive potential, and the maximum energy of the electron does not exceed $2U_p$.

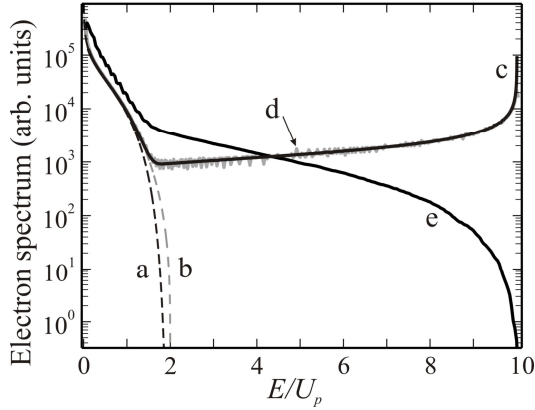

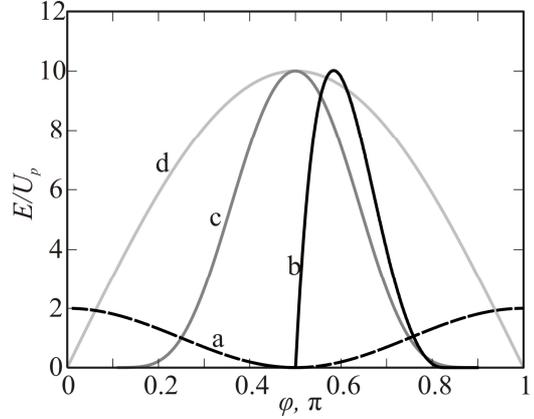

**Figure 1.** The electron spectra (a) and (b), disregarding rescattering, according to equations (11) and (12), respectively. Spectra with allowance for rescattering: (c) - calculations based on equation (18), (d) - the result of numerical simulation, (e) - calculations taking into account the quantum distribution (25).

**Figure 2.** Dependence of the final electron energy on the phase of the field: without rescattering (a) and with rescattering (b) on the metal surface. Dependence of the electron current density $J(\varphi)$ (c) and the magnitude of the field $|F(\varphi)|$ on the phase (d) (arbitrary units).

Passing from the distribution of the probability of emission in time to the distribution of electrons over the finite energy, we obtain the equation.

$$\frac{dN}{dE} = 2 \frac{dN}{dt_0} \left| \frac{dt_0}{dE} \right| \qquad (9)$$

Here $dN/dt_0 = J(t_0)$ is the electron current density of electrons emitted at a time $t_0$. From equation (9) we have

$$\left| \frac{dE}{dt_0} \right| = 4U_p \omega |\cos(\omega t_0)| \sin(\omega t_0). \qquad (10)$$

Using the relation $\cos(\omega t_0) = \sqrt{E/2U_p}$, after substituting (10) and (1) into (9), we obtain

$$\frac{dN}{dE} = \frac{1}{8\pi^2 t^2(y)} \frac{\omega}{W} \frac{\sqrt{2U_p - E}}{\sqrt{E}} \exp\left(-\frac{4W^{3/2}}{3\omega\sqrt{2U_p - E}} v(y)\right). \qquad (11)$$

Replacing in the exponential $\left(1 - \frac{E}{2U_p}\right)^{-1/2} \approx 1 + \frac{1}{2}\frac{E}{2U_p}$, we write the electron energy distribution (11) in the form

$$\frac{dN}{dE} = \frac{1}{8\pi^2 t^2(y)} \frac{\omega}{W} \frac{\sqrt{2U_p - E}}{\sqrt{E}} \exp\left(-\frac{(2W)^{3/2}}{3F_0}\left(2 + \frac{2E\omega^2}{F_0^2}\right)v(y)\right). \qquad (12)$$

Equation (12) is analogous to the distribution of electrons in the case of ionization of an atom by a low-frequency field [6], without taking into account the rescattering of electrons on the atomic core. The difference in the pre-exponential factor is due to the different types of initial electronic states. The exponent exponents in both cases are identical except for the correction factors $t(y), v(y)$, which make it possible to more accurately calculate the tunneling probability. Figure 1a, b shows the results of the spectral calculations from equations (11) and (12) for $F_0 = 0.02$ a.u., $\lambda = 3800$ nm, $W = 4.54$ eV (tungsten), and the differences are insignificant even at $E \approx 2U_p$.

**3. The electron spectrum adjusted for rescattering**
The distributions (11), (12) ignore the possibility of the electron returning to the surface of the metal with subsequent rescattering. Investigation of the related phenomenon of above-threshold ionization (ATI) of atoms has shown that only this mechanism explains the extent of the experimental electron spectra up to $10U_p$ [7]. It should be expected that rescattering is also an important mechanism for the formation of a spectrum in the emission of electrons from the metal surface in ac electric field. Let us find the change in the distribution of electrons over the finite energy due to rescattering. First, we determine the instant of time corresponding to the return of the electron to the surface of the metal. Integrating equation (6), we obtain the dependence of the electron coordinate on time

$$x(t) = \frac{F_0}{\omega}\left(\cos(\omega t_0)(t - t_0) + \frac{1}{\omega}(\sin(\omega t_0) - \sin(\omega t))\right). \tag{13}$$

The time $t_r$ the electron returns to the metal surface can be determined from the solution of the equation $x(t_r) = 0$:

$$\frac{\sin(\omega t_0) - \sin(\omega t_r)}{\omega t_0 - \omega t_r} = \cos(\omega t_0), \quad t_0 < t_r \leq \frac{5}{4}T. \tag{14}$$

We solve the equation (14) numerically, resulting in a dependence $t_r(t_0)$.

When a collision with a surface, the momentum of an electron is determined by the expression

$$p_r = \frac{F_0}{\omega}(\cos(\omega t_0) - \cos(\omega t_r)). \tag{15}$$

After the elastic reflection from the surface, the sign of the pulse changes to the opposite one, and further

$$p(t) = \frac{F_0}{\omega}(\cos(\omega t_r) - \cos(\omega t)) - p_r = \frac{F_0}{\omega}(2\cos(\omega t_r) - \cos(\omega t_0) - \cos(\omega t)). \tag{16}$$

Averaging the momentum over the period of the laser field, we find the final energy of the electron

$$E = 2U_p(2\cos(\omega t_r) - \cos(\omega t_0))^2. \tag{17}$$

In contrast to equation (8) for the energy value in the direct emission process, equation (17) gives the energy values taking into account rescattering, which can be larger $2U_p$. Numerical calculation shows that the maximum permissible finite energy of the electron upon rescattering is approximately equal $10U_p$, which agrees with the experimental spectra [8]. The dependence of energy on the phase of the field at the time of emission is shown in figure 2 a, b. To qualitatively estimate the fraction of electrons with energy in a given range, figure 2 c, d shows the dependencies $J(\varphi)$ and $|F(\varphi)|$ in arbitrary units.

To obtain the spectrum of rescattered electrons, it is necessary to take into account the electron reflection coefficient by the metal surface $R(E)$. The experimental values of the reflection coefficient depend on the orientation of the surface with respect to the crystallographic planes and the presence of an adsorbed gas on a surface [9]. It is not surprising that calculations based on simple potential models reproduce the dependence of the coefficient on reflection on energy more qualitatively than quantitatively. In the subsequent calculations we shall use the experimental values of the reflection coefficient [10].

Figure 2b shows that several values of emission moments can correspond to a single value of the final electron energy. The electron spectrum, taking into account the rescattering on the metal surface in the low-frequency limit, will have the form

$$\frac{dN}{dE} = \frac{1}{16\pi^2 t^2(y)} \frac{\omega}{W} \frac{\sqrt{2U_p - E}}{\sqrt{E}} \exp\left(-\frac{4W^{3/2}}{3\omega\sqrt{2U_p - E}} v(y)\right)\theta(E) + \\ + \sum_{i=1}^{2} R(\varepsilon(t_i)) J(t_i(E)) \left|\frac{dE(t_i(E))}{dt}\right|^{-1}. \qquad (18)$$

The first term in equation (18) describes the spectrum of electrons emitted at times corresponding to the condition $\varphi \leq 0.5$. Further motion of such electrons occurs without rescattering. The function is $\theta(E) = 1$ for $E \leq 2U_p$, and for the other values of the energy $\theta(E) = 0$. The second term in equation (18) is responsible for the spectrum of electrons that experienced rescattering on the metal surface. The summation is over roots $t_i(E)$ of equation

$$E = 2U_p (2\cos(\omega t_r(t_i)) - \cos(\omega t_i))^2, \quad T/4 < t_i(E) \leq T/2. \qquad (19)$$

Function $\varepsilon(t_i)$ is the energy of an electron at the moment of scattering on the metal surface

$$\varepsilon(t_i) = 2U_p \cos^2(\omega t_r(t_i)). \qquad (20)$$

The functions $t_r(t_i)$, $t_i(E)$, and $dE(t_i)/dt$ can be calculated only numerically, and equation (18) can not be written in an analytical form similar to equation (11). However, these dependencies can be approximated with great accuracy by polynomials:

$$\begin{aligned}
\omega t_r(t) &= a_4^r(\omega t)^4 + a_3^r(\omega t)^3 + a_2^r(\omega t)^2 + a_1^r \omega t + a_0^r, \\
\omega t_i(E) &= a_4^i(E/U_p)^4 + a_4^i(E/U_p)^3 + a_2^i(E/U_p)^2 + a_1^i(E/U_p) + a_0^i E, \\
dE(t)/dt &= U_p \omega \left(a_4(\omega t)^4 + a_3(\omega t)^3 + a_2(\omega t)^2 + a_1 \omega t + a_0\right).
\end{aligned} \qquad (21)$$

The numerical coefficients appearing in (21) are indicated in table 1.

**Table 1. Coefficients of approximations (21)**

| Functions | $a_4^k$ | $a_3^k$ | $a_2^k$ | $a_1^k$ | $a_0^k$ |
|---|---|---|---|---|---|
| $\omega t_r(t)$ | $3.8058 \times 10^1$ | $-1.2322 \times 10^2$ | $1.4908 \times 10^2$ | $-8.2033 \times 10^1$ | $1.9131 \times 10^1$ |
| $\omega t_1(E)$ | $1.8695 \times 10^{-5}$ | $-2.8458 \times 10^{-4}$ | $1.6308 \times 10^{-3}$ | $5.8890 \times 10^{-4}$ | $5.0113 \times 10^{-1}$ |
| $\omega t_2(E)$ | $1.8448 \times 10^{-5}$ | $-6.3638 \times 10^{-4}$ | $6.8915 \times 10^{-3}$ | $-4.4139 \times 10^{-2}$ | $8.0017 \times 10^{-1}$ |
| $dE(t)/dt$ | $1.4958 \times 10^4$ | $-5.0406 \times 10^4$ | $6.2940 \times 10^4$ | $-3.4434 \times 10^4$ | $6.9421 \times 10^3$ |

The results of the calculation of the spectrum with allowance for rescattering according to formula (18) are shown in figure 2c. Figure 2d shows the electron spectrum obtained for the same field and metal parameters as a result of numerical simulation.

The bend of the distribution function upwards and its sharp break in the quasistatic model are related to the classical description of the motion of an electron after its emission [11]. To eliminate this artifact, it is necessary to replace the classical electron distribution in the wave field

$$\rho(E) = \frac{1}{\pi\sqrt{E(2U_p - E)}}, \qquad (22)$$

that is implicitly present in the equations (11), (12), and (18) on the quantum analog. A quantum description of this distribution can be obtained by starting from the Schrödinger equation for an electron in a wave field at zero quasimomentum

$$i\frac{\partial \psi}{\partial t} = 2U_p \cos^2(\omega t)\psi \tag{23}$$

with the solution

$$\psi = e^{-iU_p t} \sum_{n=-\infty}^{+\infty} e^{i2n\omega t} J_n\left(-\frac{U_p}{2\omega}\right). \tag{24}$$

The energy distribution has the form $P_E \sim |J_{E/2\omega}(U_p/2\omega)|^2$. Thus, passing from the classical distribution to the quantum one, we must perform a replacement

$$\rho(E) \to |J_{E/2\omega}(U_p/2\omega)|^2, \tag{25}$$

which removes the marked energy singularity. This substitution in the second term of equation (18) formally reduces to a change in the current density

$$J(t) \to 2U_p \omega |\sin(2\omega t)| \frac{dN}{dE}, \tag{26}$$

where $dN/dE$ is calculated according to the formula (12) changed in accordance with (25). The electron spectrum obtained with allowance for (25) and (26) is shown in figure 1e.

## 4. Conclusions

We obtained an expression for the spectrum of cold emission of electrons from a metal in a strong low-frequency laser field. It is valid for small Keldysh parameters $\gamma \ll 1$ and low metal temperatures.

The calculation of the spectrum in our approach requires significantly less computational costs than the most frequently used numerical simulation for this case on the basis of two-step models [12] and includes a fundamentally important mechanism of rescattering an electron on the metal surface.